\documentclass[]{mn2e}
\usepackage{graphicx}
\usepackage{epsfig}
\usepackage{amsmath}

\newcommand       \AU           {\,{\rm AU}}
\newcommand       \cm           {\,{\rm cm}}

\newcommand       \erg          {\,{\rm erg}}
\newcommand       \eV           {\,{\rm eV}}

\newcommand       \s            {\,{\rm s}}
\newcommand       \sr           {\,{\rm sr}}

\newcommand       \GHz     {\,{\rm GHz}}

\newcommand     \gtsim  {\lower.5ex\hbox{$\buildrel > \over \sim$}}
\newcommand     \ltsim  {\lower.5ex\hbox{$\buildrel < \over \sim$}}
\newcommand     \simgt  {\lower.5ex\hbox{$\buildrel > \over \sim$}}
\newcommand     \simlt  {\lower.5ex\hbox{$\buildrel < \over \sim$}}

\newcommand       \mum          {\,{\rm \mu m}}

\newcommand       \simali       {\sim\,}

\def    \Nb	{M}
\def    \bT	{{\bf T}}

\title{
Infrared Emission of Polycyclic Aromatic
Hydrocarbon Molecules in Titan: Cyanonaphthalenes
}
\author[Zhou et al.]
{Li~Zhou$^{1,3}$\thanks{zhouli68@ncu.edu.cn},
Kaijun~Li$^{2}$\thanks{likaijun@xtu.edu.cn},
Aigen~Li$^{3}$\thanks{lia@missouri.edu},
and Zheng~Zhou$^{3,4}$\\
$^1$Department of Chemistry,
        Nanchang University,
        Nanchang 330031, China\\
$^2$Hunan Key Laboratory for Stellar and Interstellar Physics
        and School of Physics and Optoelectronics,
        Xiangtan University, Hunan 411105, China\\
$^3$Department of Physics and Astronomy,
        University of Missouri,
        Columbia, MO 65211, USA\\
$^4$John L. Miller Great Neck North High School,
        Great Neck, NY 11023, USA\\       
        }

\begin{document}

\date{}
\pagerange{\pageref{firstpage}--\pageref{lastpage}} \pubyear{2024}

\maketitle

\label{firstpage}
\begin{abstract}
As the only moon in the solar system
with a thick atmosphere,
Titan is a compelling and enigmatic world
containing a complex organic haze.
Polycyclic aromatic hydrocarbon (PAH) molecules 
are believed to play an essential role in the formation
of Titan's aerosols and haze layers.
The existence of PAHs in Titan's upper atmosphere
has been revealed by the detection of the 3.28$\mum$
emission band with Cassini’s {\it Visual and Infrared
Mapping Spectrometer} (VIMS).
However, there is little knowledge about the identity,
composition, size and abundance of PAH molecules
in Titan's atmosphere.
Due to its unprecedented sensitivity and spectral
coverage and resolution, the advent of
the {\it James Webb Space Telescope} (JWST)
could possibly enable a full characterization
of the chemical makeups of Titan's aerosols.
In particular, with a much better spectral
resolution than Cassini's VIMS, JWST's
{\it Near Infrared Spectrograph}
(and {\it Mid-Infrared Instrument})
could enable the spectral bands 
to be better resolved, potentially providing
crucial information about which PAHs are
really present in Titan's upper atmosphere.
To facilitate JWST to search for and identify
Titan's PAH molecules, we are performing
a systematic study of the photophysics of PAHs
in Titan's upper atmosphere.
As a pilot study, here we report the infrared
emission spectra of vibrationally excited
cyanonapthalenes and their ions
which are composed of two fused benzene rings
and one nitrile (–-CN) group.
The calculated emission spectra 
will help JWST to quantitatively
determine or place an upper limit on
the abundances of cyanonapthalenes
in Titan's upper atmosphere.
\end{abstract}
\begin{keywords}
ISM: dust, extinction --- ISM: lines and bands
--- ISM: molecules --- Astrochemistry
\end{keywords}

\section{Introduction}\label{sec:intro}
With an orange-brownish appearance, 
Titan, the largest moon of Saturn,
is the only known satellite
in the solar system that possesses a dense
atmosphere. 
The orange-brownish color of Titan
arises from its haze layers composed
of organic-rich aerosol particles.
While the exact composition of these aerosols
and the way they are produced are still not
fully understood, organic oligomers bearing
C, H and N atoms and aromatic materials
containing benzene rings have long been
suspected to be involved in the mechanisms
of aerosol growth (e.g., see Waite et al.\ 2007).
In particular, polycyclic aromatic hydrocarbon (PAH)
molecules are believed to be largest contributors
of Titan’s haze and have played a major role
in aerosol formation as intermediates
between small molecules and polymers
big enough to condensate as aerosols
in Titan's atmosphere (Lebonnois et al.\ 2002;
Wilson \& Atreya 2003).
Laboratory studies have shown that the inclusion
of trace amounts of aromatic species drastically
impacts the chemistry of aerosol formation
(Math\'e et al.\ 2018).


The existence of PAH molecules in Titan's atmosphere
has been supported by the detection of benzene
(C$_6$H$_6$) in its stratosphere, a single-ring
aromatic molecule and a building block of PAHs.
This detection was made by Coustenis et al.\ (2003)
based on the $\nu_4$ bending mode of benzene
at 14.85$\mum$ (674$\cm^{-1}$) observed 
by the {\it Infrared Space Observatory} (ISO).
Using the {\it Composite Infrared Spectrometer} (CIRS)
on board the Cassini spacecraft,
Vinatier et al.\ (2010) also detected the 14.85$\mum$
emission band of benzene in Titan's upper atmosphere.
Benzene has also been detected in Titan's thermosphere
by the {\it Ion and Neutral Mass Spectrometer} (INMS)
on board the Cassini spacecraft (Waite et al.\ 2007).
Using the {\it Visual and Infrared Mapping Spectrometer}
(VIMS) instrument on board Cassini, Dinelli et al.\ (2013)
obtained the near infrared (IR) spectra of Titan.
A broad, prominent emission band around 3.28$\mum$
(3049$\cm^{-1}$) was detected in Titan’s upper atmosphere.
Dinelli et al. (2013) and L\'opez-Puertas et al.\ (2013)
attributed this band to the C--H stretching emission
of PAHs of, on average, $\simali$30 carbon atoms.

In addition, laboratory simulations
have demonstrated that the formation of
PAH molecules is feasible in Titan’s atmospheric
conditions (Sagan et al.\ 1993).
Zhao et al.\ (2018) have both experimentally
and computationally shown that prototype PAHs
like anthracene  (C$_{14}$H$_{10}$)
and phenanthrene (C$_{14}$H$_{10}$)
could be synthesized via barrierless reactions
involving naphthyl radicals (C$_{10}$H$_7$\text{\textbullet})
with vinylacetylene (CH$_2$=CH–C$\equiv$CH)
in the low-temperature environments
of Titan's atmosphere.

Despite their crucial role in Titan's atmospheric chemistry
and in the formation of Titan's haze, to date, the identity,
composition, abundance, and distribution of PAH molecules
in Titan's atmosphere remain unknown
(Atreya 2007; Waite et al.\ 2010.
L\'opez-Puertas et al.\ (2013) modeled the 3.28$\mum$
emission band of Titan observed by Cassini/VIMS
and estbalished a list of 19 most abundant PAH species
(with neutral C$_{48}$H$_{22}$ and C$_{10}$H$_{8}$N
having the highest concentrations),
utilizing the {\it NASA/Ames PAH IR Spectroscopic Database}
(Bauschlicher et al.\ 2010, Boersma et al.\ 2011)
However, the identification of individual PAH molecules
is not unique or univocal.

Compared with the VIMS spectrometer on board Cassini,
the {\it James Webb Space Telescope} (JWST) has a far better
sensitity and spectral resolving power. Also thanks to
the proximity of Titan, the advent of JWST may allow us
to identify individual, specific PAH molecules
in Titan's atmosphere. To facilitate JWST to search for
and identify individual PAH molecules in Titan's upper
atmosphere, we launch a systematic exploration of
the vibrational excitation and radiative de-excitation
of a large number of specific PAH molecules in Titan.
For demonstrative purpose, in this paper we present
a pilot study, with cyanonapthalene as our target molecules.
We note that the purpose of this work
is not to model the Titan's IR emission, instead,
it is just a pilot study with cyanonapthalenes
selected for demonstration.
This paper is organized as follows.
We briefly summarize the physics underlying
the vibrational excitation and de-excitation
of cyanonaphthalenes in \S\ref{sec:model}.
The IR emission spectra of cyanonaphthalenes
are presented and discussed in \S\ref{sec:results}
and summarized in \S\ref{sec:summary}. 





\begin{figure}
\centering
\includegraphics[height=8cm]{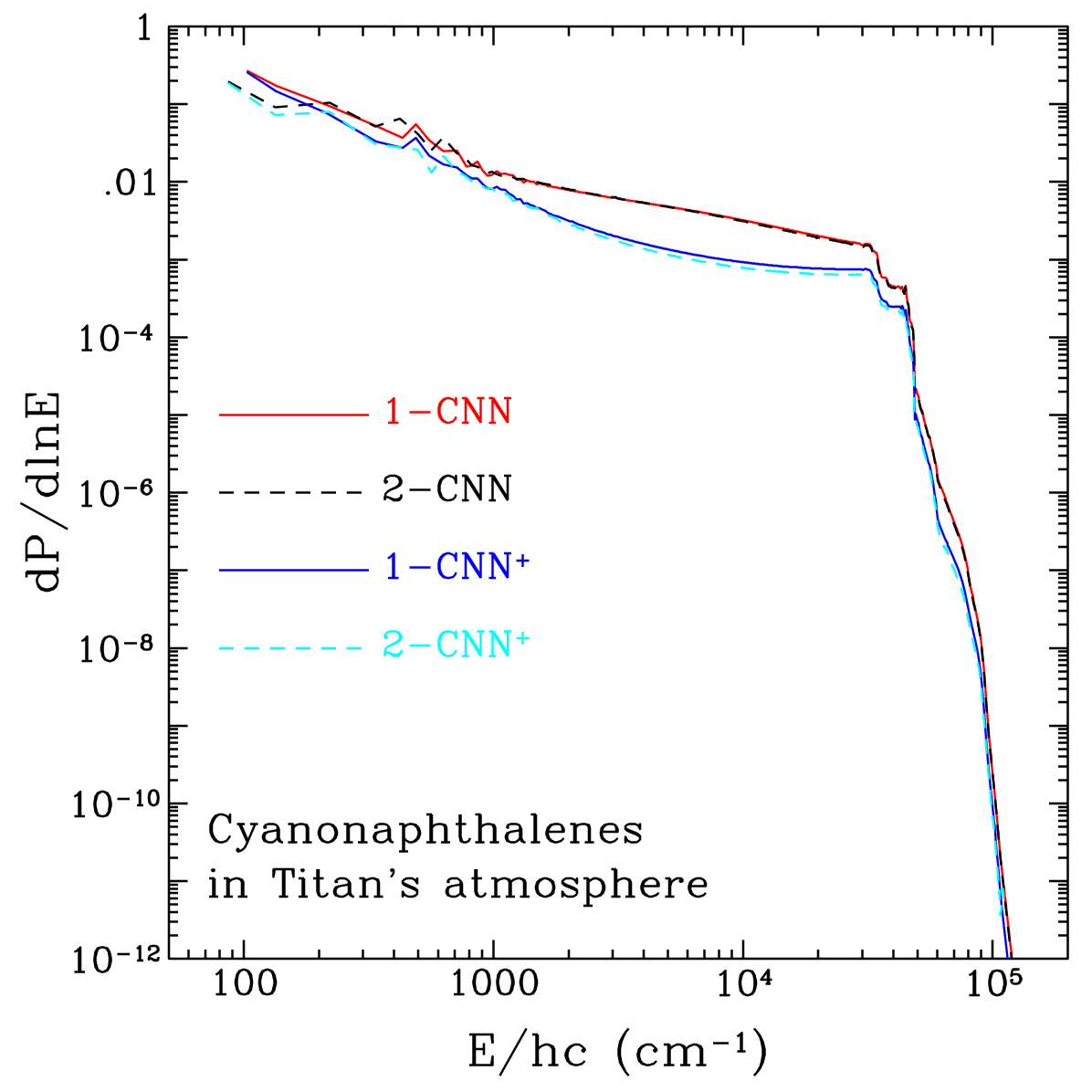}
\caption{
       \label{fig:dPdlnE}
        Vibrational energy probability distribution 
        functions for the excited vibrational states
        of 1-CNN, exposed to the solar radiation
        in Titan's upper atmosphere.
         }
\end{figure}

\section{Photo-excitation of Cyanonaphthalenes
           in Titan's Atmosphere}\label{sec:model}
Cyanonapthalenes (C$_{10}$H$_7$CN) consist of
two fused benzene rings and substitute a nitrile
(–-CN) group for a hydrogen atom.
For demonstrative purpose, we select
cyanonapthalenes for the present study
because they are the very first specific PAH
molecules ever identified in the interstellar medium (ISM).
Using the 100\,m Green Bank Telescope (GBT),
McGuire et al.\ (2021) conducted radio observations
in the frequency range of 8 to 34$\GHz$,
of the dark molecular cloud TMC-1
located within the Taurus Molecular Cloud.
They reported the detection of
the rotational transitions of
1-cyanonaphthalene (1-CNN)
and 2-cyanonaphthalene (2-CNN),
two isomers of cyanonaphthalene.
Although there is no apparent link
between the ISM and Titan's atmosphere,
the detection of interstellar cyanonaphthalenes
at least shows that cyanonaphthalenes
can form in extraterrestrial environments. 
%





We select cyanonapthalenes also because
in the N-rich atmosphere of Titan,
PAH molecules are likely to contain N atoms.
N$_2$ and methane (CH$_4$) are the predominant
species in Titan's atmosphere
(e.g., see Cui et al.\ 2009, 2012, 2016).
Bombarded by energetic particles
from Saturn's magnetosphere and/or
irradiated by solar ultraviolet (UV) photons,
both N$_2$ and CH$_4$ could dissociate into
radicals and ions. This triggers a complex organic
chemistry in the upper atmosphere and impacts
the formation and composition of PAHs and aerosol
(e.g., the numerous C, H and N compounds
in Titan's atmosphere could lead to the formation
of  nitrogenous PAHs; Waite et al.\ 2007).

In principle, PAHs in Titan's atmosphere could
be excited by solar photons and energetic particles
from Saturn’s magnetosphere.
However, the fact that the 3.28$\mum$ emission
detected by Cassini/VIMS was only seen during
daytime and vanished at night clearly demonstrates
that the PAH molecules in Titan's atmosphere
are definitely pumped by solar radiation.

To characterize how cyanonapthalenes absorb
solar photons, we adopt the UV absorption
cross sections of cyanonapthalenes synthesized
by Li et al.\ (2023).
For the IR absorption cross sections which
determine how cyanonaphthalenes emit,
we take the vibrational frequencies and intensities
calculated by Bauschlicher (1998)
from the B3LYP density functional theory
in conjunction with the 4-31G basis set
which are available from
the {\it NASA Ames PAH IR Spectroscopic Database}
(Boersma et al.\ 2014, Bauschlicher et al.\ 2018, 
Mattioda et al.\ 2020).
Cyanonaphthalenes have 19 atoms and,
under the harmonic oscillator approximation,
51 vibrational modes (lines).
Following Allamandolla et al.\ (1999)
and Li et al.\ (2023), we represent each vibrational line
by a Drude function, characterized by the peak wavelength
and intensity of the vibrational transition.
In addition, we assign a width of 30$\cm^{-1}$
for each line, consistent with the natural line width
expected from a vibrationally excited PAH molecule 
(see Allamandola et al.\ 1999).

Upon absorption of a solar photon,
cyanonaphthalenes will undergo stochastic
heating since their energy contents are often
smaller than the energy of a single solar photon.  
We model the stochastic heating 
of cyanonaphthalenes by employing
the ``exact-statistical'' method
of Draine \& Li (2001).
We characterize the state of
a cyanonaphthalene molecule
(i.e., 1-CNN or 2-CNN)
by its vibrational energy $E$, 
and group its energy levels into 
$(\Nb+1)$ ``bins'',
where the $j$-th bin ($j$\,=\,0,\,...,\,$\Nb$) 
is $[E_{j,\min},E_{j,\max})$, 
with representative energy 
$E_j$\,$\equiv$\,$(E_{j,\min}$+$E_{j,\max})$/2,
and width 
$\Delta E_j$\,$\equiv$\,$(E_{j,\max}$--$E_{j,\min})$.
Let $P_j$ be the probability of finding 
1-CNN (or 2-CNN) in bin $j$ with energy $E_j$.
The probability vector $P_j$ evolves 
according to
\begin{equation}
dP_i/dt = \sum_{j\neq i} \bT_{ij} P_j 
- \sum_{j\neq i} \bT_{ji}P_i ~~,~~ i\,=\,0,...,\Nb ~~,
\end{equation}
where the transition matrix element $\bT_{ij}$ is
the probability per unit time for
1-CNN (or 2-CNN) in bin $j$ 
to make a transition to one of the levels in bin $i$. 
We solve the steady state equations
\begin{equation}\label{eq:steadystate}
\sum_{j\neq i} \bT_{ij} P_j
= \sum_{j\neq i} \bT_{ji}P_i ~~,~~ i\,=\,0,...,\Nb ~~
\end{equation}
to obtain the $\Nb$+1 elements of $P_j$,
and then calculate the resulting IR emission spectrum
(see eq.\,55 of Draine \& Li 2001).

In calculating the state-to-state transition rates
$\bT_{ji}$ for transitions $i$$\rightarrow$$j$,
we first calculate the excitation rates
$\bT_{ji}$ (i.e., $i$\,$<$\,$j$), which are simply
the photon absorption rates. The deexcitation rates
$\bT_{ij}$ can be determined from the excitation rates
$\bT_{ji}$ based on the detailed balance analysis
of the Einstein $A$ coefficient.
This requires the knowledge of
the degeneracies $g_i$ and $g_j$
of bin $i$ and bin $j$, which are
the numbers of energy states 
in bins $i$ and $j$, respectively.
Based on the frequencies of 
all the 51 vibrational modes
of 1-CNN (and 2-CNN)
calculated by Bauschlicher (1998), 
we employ the Beyer-Swinehart numerical algorithm
(Beyer \& Swinehart 1973, Stein \& Rabinovitch 1973)
to calculate the vibrational density of states 
and therefore the degeneracies 
for each vibrational energy bin.
We finally solve the steady-state state probability 
evolution equation (see eq.\,\ref{eq:steadystate})
and then calculate the resulting IR emission spectrum.
For computational convenience, 
we consider 500 energy bins
(i.e., $M=500$).
%

\begin{figure*}
\begin{center}
\hspace{-1cm}
\begin{minipage}[t]{0.4\textwidth}
\resizebox{8.2cm}{7.5cm}{\includegraphics[clip]{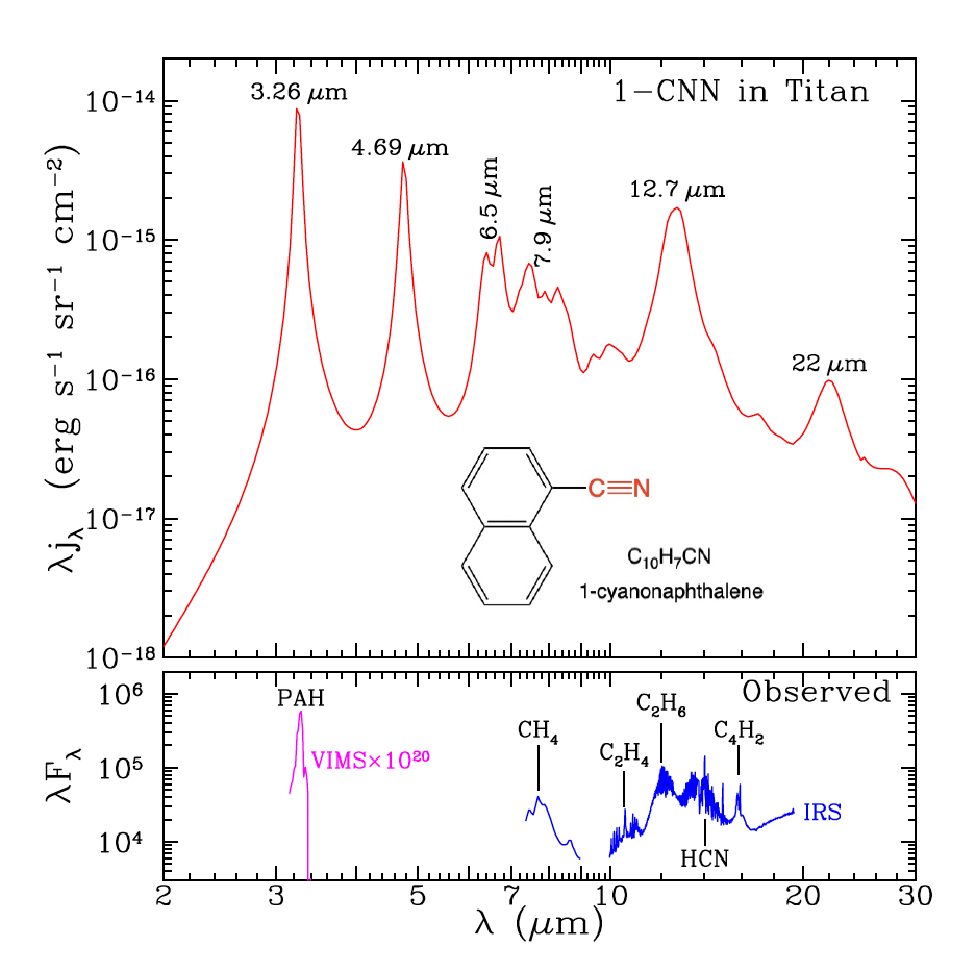}}\vspace{-0.5cm}
\end{minipage}
\hspace{1.5cm}
\begin{minipage}[t]{0.4\textwidth}
\resizebox{8.2cm}{7.5cm}{\includegraphics[clip]{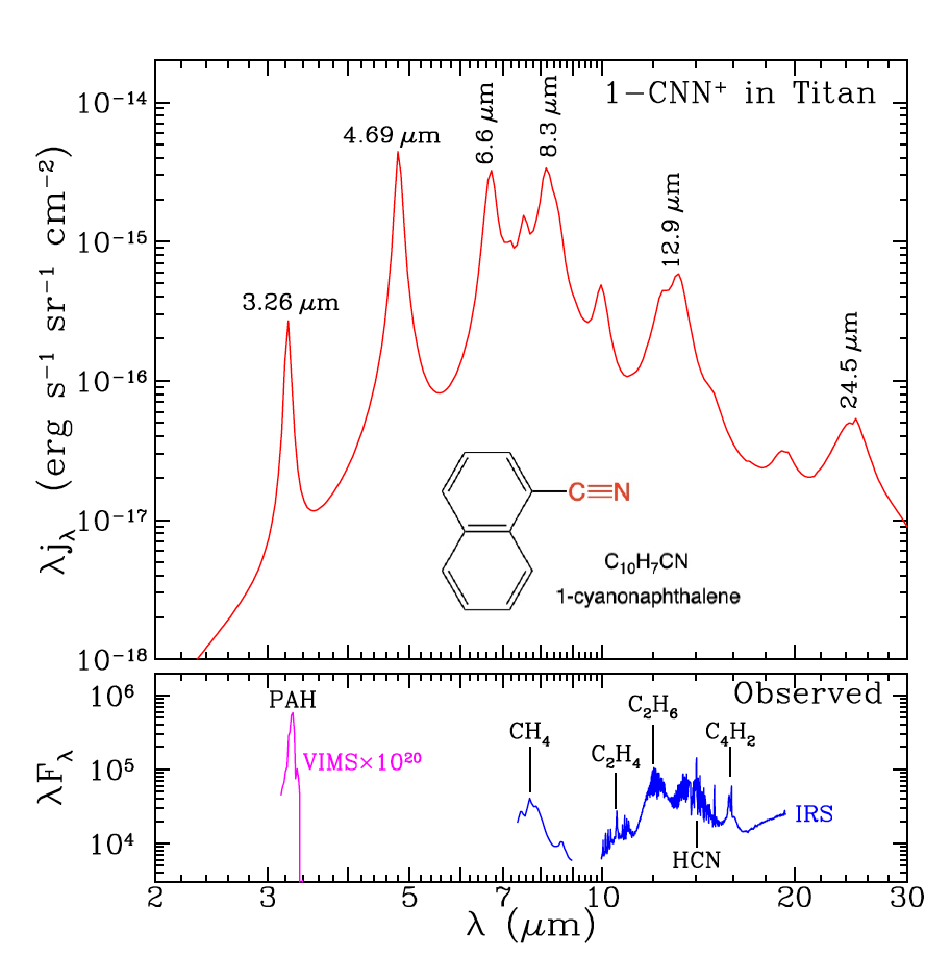}}\vspace{-0.5cm}
\end{minipage}
\end{center}
\caption{
         \label{fig:cnn1}
         Top left panel (a): IR emission spectrum of 1-CNN
         excited by solar photons in Titan's upper atmosphere.
         Top right panel (b):  Same as Left panel (a) but for
         ionized 1-CNN.
         For comparison, we show in the bottom panels
         the Cassini/VIMS spectrum of Titan's upper atmosphere
         at 3.15--3.40$\mum$ (with the $\nu_3$ emission of
         CH$_4$ at 3.3$\mum$ subtracted), as well as
         the {\it Spitzer}/IRS spectra obtained with
         the Short Wavelength-Low Resolution module
         at 7.36--8.97$\mum$ and the Short Wavelength-High
         Resolution module at 9.96-19.35$\mum$.
         The unit for the Y-axis ($\lambda F_\lambda$)
         of the bottom panels is $\erg\s^{-1}\sr^{-1}\cm^{-2}$.
         }
\end{figure*}

\section{Results and Discussion}\label{sec:results}
At a heliocentric distance of $r_h=9.5\AU$,
cyanonapthalenes in the upper atmosphere of Titan
is vibrationally excited by the solar photospheric
radiation diluted by a factor of
$\left(R_\odot/2r_h\right)^2$,
where $R_\odot$ is the solar radius.
The mean energy of the solar photons
absorbed by cyanonapthalenes is
$\langle h\nu\rangle_{\rm abs}\approx4.6\eV$.
As we do not distinguish the UV absorption
properties of 1-CNN from that of 2-CNN
and their ions, the mean photon energies are
essentially the same for 1-CNN and 2-CNN
and their ions. In principle, a somewhat lower
mean photon energy $\langle h\nu\rangle_{\rm abs}$
is expected for ions since, upon ionization,
their absorption edges shift to longer
wavelengths (see Li \& Draine 2002).
The electronic structures of 1- and 2-CNN are
not identical and therefore their UV absorption
properties should also be somewhat different.
However, as we do not have any UV experimental
data for 2-CNN and cyanonapthalene cations,
we do not distingish their UV absorption cross sections. 

We show in Figure~\ref{fig:dPdlnE}
the energy probability distribution functions
of 1- and 2-CNN and their cations.
We see that, upon the absorption of a solar photon,
due to their small heat contents, cyanonapthalenes
are excited to high vibrational states
(with high vibrational energies).
Subsequently, they cool down rapdily
by emitting IR photons and then spend
most of the time staying in low vibrational states
until they encounter another solar photon. 
As the adopted UV absorption properties
are identical for 1- and 2-CNN and their cations,
their energy probability distribution functions
do not differ much from each other.
The small differences among 1- and 2-CNN
and their cations seen in Figure~\ref{fig:dPdlnE}
arise from their different IR emission properties
(e.g., neutral cyanonapthalenes emit strongly
at 3.26$\mum$ while their cations emit strongly
at $\simali$6--8$\mum$).

Figure~\ref{fig:cnn1}a shows the IR emissivity
(erg$\s^{-1}\sr^{-1}\cm^{-1}$) per molecule
expected for 1-CNN in Titan's upper atmosphere.
It is apparent that 1-CNN would emit several
pronounced spectral bands: a C--H stretching band
at 3.26$\mum$, a C--N stretching band
at 4.69$\mum$, and a broad complex
at $\simali$6--9$\mum$ consisting of
a number of sub-features
arising from C--C stretching
and C--H in-plane bending vibrations,
as well as a C--H out-of-plane bending
band at 12.7$\mum$.
In addition, 1-CNN also exhibits
a C-C-C skeletal bending band
at $\simali$22$\mum$.
While the 6--9$\mum$ wavelenth range
is generally ``crowded'' for PAH molecules
(e.g., in many astrophysical environments
PAH molecules collectively emit at a distinctive
set of bands at 6.2, 7.7 and 8.6$\mum$; see Li 2020),
the bands at 3.26, 4.69, 12.7 and 22$\mum$
are rather characteristics of 1-CNN and,
particularly, their relative intensities are diagnostic
of the presence and abundance of 1-CNN in Titan.

We have also calculated the IR emissivity
(per molecule) for 1-CNN cation
in Titan's upper atmosphere.
As shown in Figure~\ref{fig:cnn1}b,
compared to its neutral counterpart,
1-CNN cation emits much less at
the 3.26$\mum$ C--H stretching band
and the 12.9$\mum$ C--H out-of-plane
bending bnd; on the other hand, 1-CNN cation
emits more strongly around 6--9$\mum$.
It is interesting to note that, upon ionization,
there is not much change on the 4.69$\mum$
C--N stretching emission band. 

\begin{figure*}
\begin{center}e
\hspace{-1cm}
\begin{minipage}[t]{0.4\textwidth}
\resizebox{8.2cm}{7.5cm}{\includegraphics[clip]{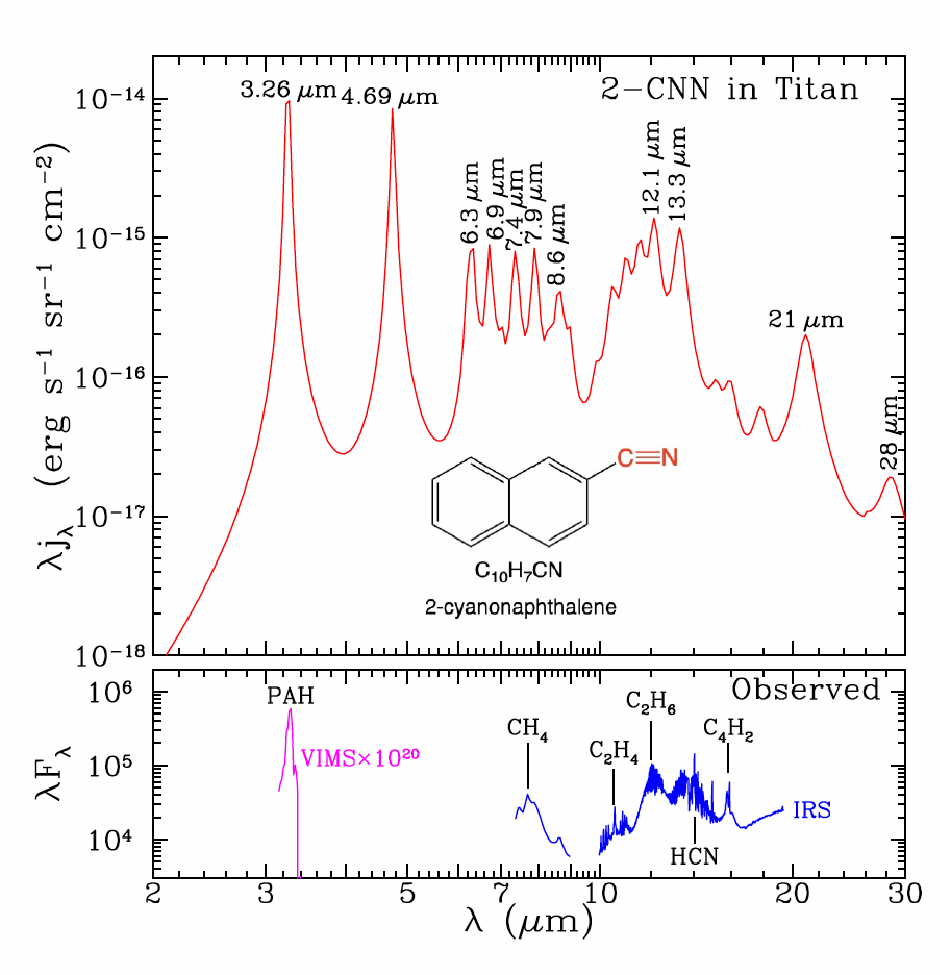}}\vspace{-0.5cm}
\end{minipage}
\hspace{1.5cm}
\begin{minipage}[t]{0.4\textwidth}
\resizebox{8.2cm}{7.5cm}{\includegraphics[clip]{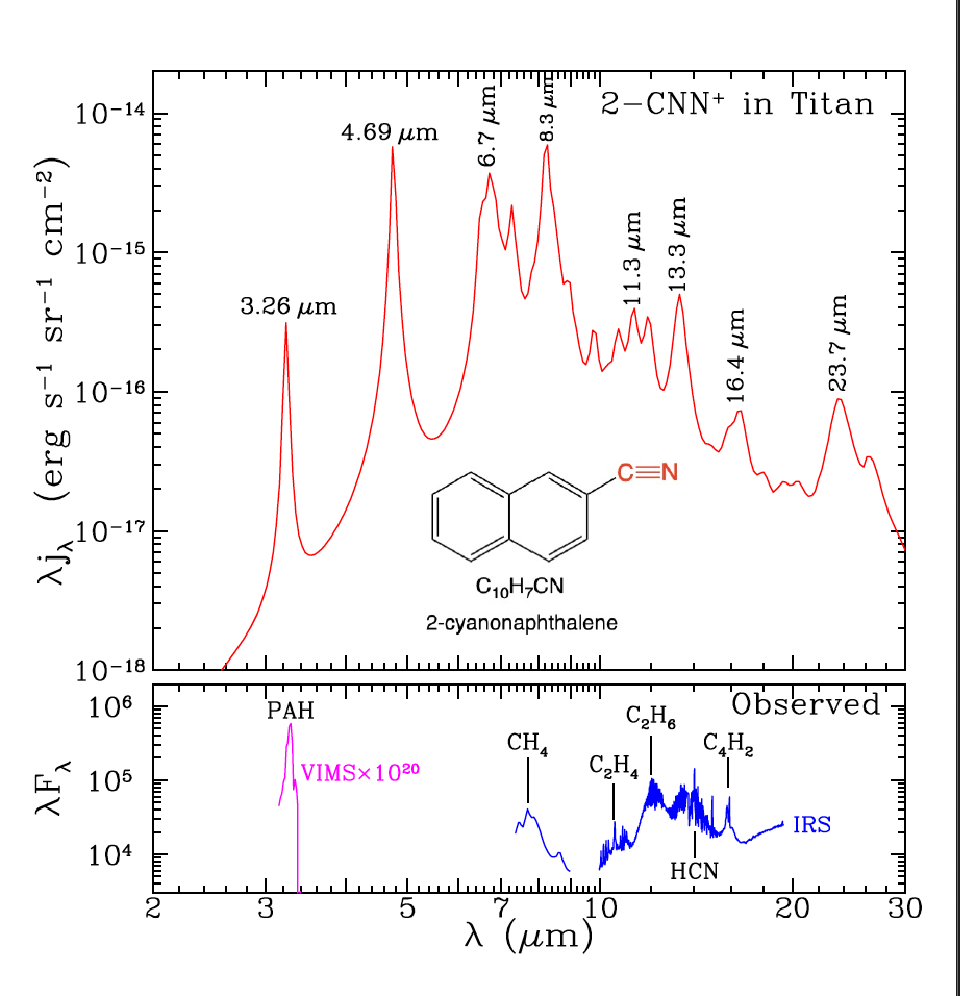}}\vspace{-0.5cm}
\end{minipage}
\end{center}
\caption{
         \label{fig:cnn2}
         Same as Figure~\ref{fig:cnn1} but for 2-CNN
         and its cation.
         }
\end{figure*}

Figure~\ref{fig:cnn2}a presents the IR emission
spectrum computed for 2-CNN
in Titan's atmosphere. 
A first glance of Figure~\ref{fig:cnn2}a
reveals that the IR emission spectrum of
2-CNN closely resembles that of 1-CNN:
2-CNN also exhibits a prominent 3.26$\mum$
band attributed to C--H stretch and a prominent
4.69$\mum$ band attributed to C--N stretch.
The major difference between 1-CNN and 2-CNN
is that, while the C--C stretching bands at
$\simali$6--9$\mum$ and C--H out-of-plane
bending bands at $\simali$11--14$\mum$
of 2-CNN show a number of sub-features,
for 1-CNN these sub-features coalesce into
three broad bands peaking at 6.5, 7.9 and 12.7$\mum$.
Also, the C-C-C skeletal bending band
of 2-CNN occurs at a somewhat shorter
wavelength of $\simali$21$\mum$
($\simali$22$\mum$ for 1-CNN).

We have also computed the IR emission
spectrum of 2-CNN cation.
As shown in Figure~\ref{fig:cnn2}b,
it is closely similar to that of 1-CNN cation.
The major difference is that,
the C--H out-of-plane stretching bands
of 1-CNN cation fall into one single broad band,
2-CNN cation shows two relatively narrow
bands peaking at $\simali$11.3 and 13.3$\mum$.
In addition, 2-CNN cation shows an emission band
at 16.4$\mum$ which is not seen in 1-CNN cation.
Finally, the C-C-C skeletal bending band
of 2-CNN occurs at $\simali$23.7$\mum$,
while 1-CNN cation peaks at $\simali$24.5$\mum$.

While observationally the identification of
cyanonaphthalenes through their characteristic
IR vibrational bands may be complicated
by the fact that many other PAH species
also emit a rich set of C--H and C--C
stretching and bending bands in the IR,
the detection of the 3.26 and 4.69$\mum$ bands,
and, to a less degree, the 21 or 24$\mum$ band,
could {\it potentially} allow one to identify
cyanonaphthalenes in Titan's atmosphere.
For comparison, we show in
Figures~\ref{fig:cnn1},\,\ref{fig:cnn2}
the Cassini/VIMS spectrum of Titan
at $\simali$3.15--3.40$\mum$, 
after the $\nu_3$ emission of CH$_4$
was subtracted (Dinelli et al.\ 2013).
With a spectral resolution of $\simali$16\,nm, 
the IR channel (0.85–-5.2$\mum$)
of the Cassini/VIMS imaging spectrometer
clearly detected the $R$, $Q$, and $P$ branches
of the 3.3$\mum$ emission bands of CH$_4$
in Titan’s upper atmosphere
(Garc\'ia-Comas et al.\ 2011).
Dinelli et al.\ (2013) found that CH$_4$
emission can explain very well the measured
Cassini/VIMS spectrum at wavelengths longer
than 3.3$\mum$, but it clearly underestimates
the emission at wavelengths shorter than 3.3$\mum$.
With the simulated CH$_4$ emission subtracted
from the measured Cassini/VIMS spectrum,
the residual spectrum shows a clear peak
around 3.28$\mum$, characteristic of
the C--H stretches of small PAHs.
In comparison, the C--H stretches
of cyanonaphthalenes and their cations
peak around 3.26$\mum$.

Although the wavelength span of
Cassini/VIMS's IR channel extends
up to 5.2$\mum$, to our knowledge,
no quality spectrum had been reported
at wavelengths longer than $\simali$3.4$\mum$.
Therefore, it is not possibe to tell
if the C--N stretches of cyanonaphthalenes
and their cations at 4.69$\mum$ were present
(or absent) in the Cassini/VIMS spectrum.
Also, the {\it Infrared Spectrograph}
(IRS) on board the {\it Spitzer Space Telescope}
lacks coverage in the 3--5$\mum$ area of interest.
On the other hand, the {\it Near Infrared Spectrograph}
(NIRSpec) instrument on board JWST has a much better
spectral resolution than Cassini/VIMS,
e.g., the {\it Integral Field Unit} (IFU)  mode
of JWST/NIRSpec provides a spectral resolution
up to $\simali$2,700 in the 2.9--5.0$\mum$
(Filter F290LP) spectral range (Nixon et al.\ 2016).
This could potentially enable the 3.3$\mum$ CH$_4$
and 3.28$\mum$ PAH bands to be better resolved.
This could also potentially enable the detection
of the 4.69$\mum$ C--N stretching band and
provides crucial information about which PAHs
are really present in Titan's upper atmosphere.


We admit that it is not clear
if the detection of the 3.26$\mum$ C--H 
and 4.69$\mum$ C--N stretching bands
could {\it uniquely} pinpoint the presence
of cyanonaphthalenes,
as other cyano-substituted PAHs
may also emit at similar wavelengths.
Indeed, as shown in Figures~\ref{fig:cnn1},\,\ref{fig:cnn2},
it is hard to compare a single PAH molecule with the entire
observed spectrum.
To this end, a systematic calculation of
the IR emission spectra of these molecules
would be crucial. One can imagine that
different-sized cyano-containing aromatic molecules
would exhibit different C--H/C--N band ratios
since, with different energy contents,
they are expected to be excited to different
energy levels by the same solar photon.
Without a full exploration of a large number
of PAH species, it is difficult to assign 
the fingerprints of each specific PAH molecule
and to decipher the IR emission spectrum of
Titan obtained by JWST.


Figures~\ref{fig:cnn1},\,\ref{fig:cnn2} also
show the {\it Spitzer}/IRS spectrum of Titan
at $\simali$7.36--8.97$\mum$ and
$\simali$9.96--19.35$\mum$ obtained by
Coy et al.\ (2023) in the 2004--2009 time period.
It is apparent that CH$_4$, HCN, HC$_3$N
and various simple hydrocarbon molecules
such as C$_2$H$_2$, C$_2$H$_4$, C$_2$H$_6$,
C$_3$H$_4$, C$_3$H$_6$, C$_3$H$_8$,
and C$_4$H$_2$ are clearly visible
in the {\it Spitzer}/IRS spectrum of Titan
(see Coy et al.\ 2023).
In these wavelength ranges, interstellar PAHs
exhibit a distinctive set of pronounced emission
bands at 6.2, 7.7, 8.6, 11.3 and 12.7$\mum$
(see Li 2020). To examine whether such PAHs
are present in Titan, careful subtraction of
the emission bands from CH$_4$ and various
hydrocarbon and nitrile molecules from
the {\it Spitzer}/IRS spetcrum is essential
(e.g., the $\nu_4$ band of CH$_4$ at 7.7$\mum$
would complicate the detection of the 7.7$\mum$
band of PAHs, and the 10.6$\mum$ band of
C$_2$H$_4$ as well as the 12.3$\mum$ band of
C$_2$H$_6$ would complicate the detection
of the 11.3$\mum$ band of PAHs).
Particularly, to explore the possible presence of
individual specific PAH molecules in Titan,
one requires high spectral resolution and
high signal-to-noise (S/N) data.
These molecules exhibit many emission bands
in the 5--28$\mum$ wavelength range
fully accessible to JWST's {\it Mid-Infrared Instrument}
(MIRI) that may be blended at the lower resolving power
of {\it Spitzer}/IRS ($R$\,$\simali$60--127 for its
Short Wavelength-Low Resolution module
at 5.1--14.3$\mum$, and $R$\,$\simali$600
for its Short Wavelength-High Resolution module
at 9.9--19.5$\mum$), but may show up 
in the data obtained with JWST/MIRI's medium-resolution
spectrometer (MRS) of a spectral resolving power of
$R\simali$1,500--3,500.
In addition to the increased resolving power,
the much increased sensitivity of JWST/MIRI will
yield improved detections of weak emission
bands and better separation of bands that
are mixed together at lower resolutions.
We note that, Titan has recently been observed
by JWST's NIRSpec and MIRI instruments
and the analysis of these data is ongoing
(Nixon et al.\ 2023, Teanby et al.\ 2023).
The model emission spectra of specific
PAH molecules like that presented here
lay the groundwork for searching for and
identifying PAH molecules in JWST's NIRSpec and MIRI
spectra and fill the gap in the current
spectroscopic knowledge of specific PAHs
and that expected to be present in Titan,
and therefore maximize the scientific products
from JWST observations.
%

\section{Summary}\label{sec:summary}
As a pilot exploration of the photophysics
of PAH molecules in Titan's upper atmosphere,
we have modeled the IR emission of two isomers
of cyanonapthalenes and their ions
vibrationally excited by solar photons.
It is found that with prominent bands
occuring at 3.26$\mum$ (C--H stretch)
and 4.69$\mum$ (C--N stretch),
cyanonapthalenes may be detectable
in Titan's upper atmosphere
by the NIRSpec spectrographt on board JWST.
Cyanonapthalenes also exhibit several
characteristic bands at $\simali$6--9$\mum$
and $\simali$10.5--13.5$\mum$.
But these bands are less effective in
diagnosing cyanonapthalenes
since they may be mixed with
that of other aromatic molecules.
The model emission spectra of
a large number of specific PAH molecules
like that of cyanonapthalenes reported here
lay the groundwork for searching for and
identifying specific PAH molecules in
the high spectral resolution and high S/N ratio
spectra of Titan obtained with JWST's NIRSpec
and MIRI instruments.

\section*{Acknowledgements}
We thank B.M.~Broderick, B.P.~Coy,
B.T.~Draine, I.~Mann, B.A.~McGuire,
C.E.~Mentzer, C.A.~Nixon,
E.F.~van Dishoeck, B.~Yang
and the anonymous referee
for stimulating discussions and suggestions.
We thank B.M.~Dinelli and M.~L\'opez Puertas
for providing us with their Cassini/VIMS data
and B.P.~Coy and C.A.~Nixon for providing
their {\it Spitzer}/IRS data of Titan.
LZ is supported by NSFC~12363008.
KJL is supported by
the National Key R\&D Program of China
under No.\,2017YFA0402600,
and the NSFC grants 11890692,
12133008, and 12221003,
as well as CMS-CSST-2021-A04.

\section*{Data Availability}
The data underlying this article will be shared
on reasonable request to the corresponding authors.


\end{document}